%% file: techrep.tex
\documentclass[10pt,a4paper,onecolumn]{article}

\usepackage[ruled]{algorithm}
\usepackage{algpseudocode}
\usepackage[isolatin]{inputenc}
\usepackage[OT1]{fontenc}
\usepackage{wide}
\usepackage{multicol}
\usepackage{url}
\usepackage{epsfig}
\usepackage{latexsym}

\newcommand{\BEGLIST}{\begin{list}{}{\partopsep -0pt \parsep -0pt \listparindent 0pt \labelwidth .5in}}
\newcommand{\ENDLIST}{\end{list}}

\alglanguage{pseudocode}
\algblockdefx{RepeatForever}{EndRepeatForever}{{\bf repeat forever}}{{\bf end repeat}}

\algblockdefx{Begin}{End}{{\bf begin}}{{\bf end}}
\algblockdefx{While}{EndWhile}[1]{{\bf while}\ {#1}\ {\bf do}}{{\bf end while}}

\newcommand{\send}[2]{{\bf send}$\langle #1 \rangle$\ {\bf to}\ $#2$}
\newcommand{\receive}[2]{{\bf receive}$\langle #1 \rangle$\ {\bf from}\ $#2$}
\newcommand{\token}{\mathtt{ResT}}
\newcommand{\pusher}{\mathtt{PushT}}
\newcommand{\priority}{\mathtt{PrioT}}
\newcommand{\ctrl}{\mathtt{ctrl}}
\newcommand{\hasprio}{\mathtt{Prio}}
\newcommand{\state}{\mathtt{State}}
\newcommand{\rset}{\mathtt{RSet}}
\newcommand{\request}{\emph{Req}}
\newcommand{\inside}{\emph{In}}
\newcommand{\out}{\emph{Out}}
\newcommand{\enter}{\mathtt{EnterCS}()}
\newcommand{\release}{\mathtt{ReleaseCS}()}
\newcommand{\need}{\mathtt{Need}}
\newcommand{\next}{\mathtt{Succ}}
\newcommand{\myC}{\mathtt{myC}}

\newcommand{\K}{\mathtt{k}}
\newcommand{\elle}{\ell}
\newcommand{\klexclu}{$\K$\mbox{-}out\mbox{-}of\mbox{-}$\elle$\ exclusion}
\newcommand{\stoken}{\mathtt{SToken}}
\newcommand{\ok}{\mathtt{Ok}}
\newcommand{\sprio}{\mathtt{SPrio}}
\newcommand{\spusher}{\mathtt{SPush}}
\newcommand{\reset}{\mathtt{Reset}}
\newcommand{\timeout}{\mathtt{TimeOut}()}
\newcommand{\restarttimer}{\mathtt{RestartTimer}()}
\newcommand{\CMAX}{\mathtt C_{\mathtt{MAX}}}
\newcommand\rmp[1]{}
\newcommand{\Root}{\emph{r}}

\newcommand{\valu}{\mbox{\it value\/}}
\newcommand{\type}{\mbox{\it type\/}}

\newtheorem{definition}{Definition}
\newtheorem{theorem}{Theorem}
\newtheorem{lemma}{Lemma}
\newtheorem{remark}{Remark}

\newenvironment{proof}{{\bf Proof. } }{{\hfill $\Box$}\vspace{.5pc}}

\author{Ajoy K. Datta$^1$ \and St\'ephane Devismes$^2$ \and Florian Horn$^3$ \and Lawrence L. Larmore$^1$}

\title{Self-Stabilizing $\K$\mbox{-}out\mbox{-}of\mbox{-}$\elle$\ Exclusion on Tree Networks}

\date{}

\begin{document}

\maketitle

\begin{center}
$^1$School of Computer Science, University of Nevada Las Vegas, \url{{datta,larmore}@cs.unlv.edu}\\
$^2$VERIMAG, Universit\'e Grenoble 1 Joseph Fourier, \url{stephane.devismes@imag.fr}\\
$^3$LIAFA, Universit\'e Paris 7 Denis Diderot, \url{horn@liafa.jussieu.fr}
\end{center}

\input{abstract}

\input{intro}

\input{model}

\input{algo}

\input{preuve}

\input{ccl}

\bibliographystyle{plain} 
\bibliography{biblio}

\end{document}

%% file: abstract.tex
\begin{abstract}
{\small
In this paper, we address the problem of \klexclu,
a generalization of the mutual exclusion problem,
in which there are $\elle$ units of a shared resource, and any process
can request up to $\K$ units ($1\leq\K\leq\elle$).
We propose the first deterministic self-stabilizing distributed
\klexclu\  protocol in message-passing systems for
asynchronous oriented tree networks which assumes bounded
local memory for each process.

\paragraph{Keywords:}
Fault-tolerance, self-stabilization, resource allocation, \klexclu,
oriented tree networks.}
\end{abstract}

%% file: intro.tex
\section{Introduction}\label{sect:intro}

The basic problem in resource allocation is the management of shared resources,
such as printers or shared variables. The use of such resources by an agent
affects their availability for the other users. In the aforementioned cases, 
at most one agent can access the resource at any
time, using a special section of code called a \emph{critical section}.
The associated protocols must guarantee the \emph{mutual exclusion}
property~\cite{R86b}: the critical section can be executed by at most one
process at any time. The $\elle$\mbox{-}exclusion
property~\cite{FLBB89j} is a
generalization of mutual exclusion, where $\elle$ processes can execute the
critical section simultaneously. Thus, in $\elle$\mbox{-}exclusion, $\elle$
units of a same resource ({\em e.g.}, a pool of IP addresses) can be
allocated. This problem can be generalized still further by considering
heterogeneous requests, {\em e.g.}, bandwidth for audio or video streaming.
The {\klexclu} property~\cite{R91c} allows us to deal with such requests;
requests may vary from
1 to $\K$ units of a given resource, where $1 \leq \K \leq \elle$.

\vspace*{-10pt}

\paragraph{Contributions.} In this paper, we propose a (deterministic)
\emph{self-stabilizing} distributed \klexclu\
 protocol
for asynchronous oriented tree networks. 
A protocol is self-stabilizing~\cite{Dij74}
if, after transient faults hit the system and
place it in some arbitrary global state, the systems recovers from this
catastrophic situation without external (\emph{e.g.} human) intervention in
finite time.
Our protocol is written in
the message-passing model, and assumes bounded memory per process.
To the best of our knowledge, there is no prior protocol of
this type in the literature.

Obtaining a self-stabilizing solution for the \klexclu\ problem 
in oriented trees
 is desirable, but also complex.
Our main reason for dealing with oriented trees is that
extension to general rooted networks is
trivial;
it consists of running the protocol concurrently with a
spanning tree construction (for message passing systems),
such as given in~\cite{AB97j,DDT05c}.
In the other hand, the complexity of the solution
comes from the fact that the problem is a generalization of
mutual exclusion. This is exacerbated by the difficulty of obtaining
self-stabilizing solutions in message-passing system (the more realistic
model), as underlined by the impossibility result of Gouda and
Multari~\cite{GM91j}.

Designing protocols for such problems on realistic
systems often leads to obfuscated solutions. A direct consequence is then
the difficulty of checking, or analyzing the solution. To circumvent this
problem, we propose, here, a step-by-step approach.
We start from a ``naive'' non-operating circulation of $\elle$ resource tokens.
Incrementally, we augment this solution with several other types of
tokens until we obtain a
correct non fault-tolerant solution.  We then introduce an additional control
mechanism that guarantees self-stabilization assuming unbounded local
memory.  Finally, we modify the protocol to accommodate bounded local memory.
We validate our approach by showing correctness
and analyzing waiting time, a crucial parameter in resource allocation.

\paragraph{Related Work.} Two kinds of protocols
are widely used in the literature to solve the \klexclu\ problem: 
permission-based protocols, and $\elle$-token circulation.
All non self-stabili\-zing solutions currently in the literature are
permission-based. In a permission-based protocol,
any process can access a resource after receiving permissions from all
processes \cite{R91c}, or from
the processes constituting
its quorum
\cite{MBRA98j,MT99c}.
There exist two self-stabilizing solutions
for \klexclu\ on the
oriented rooted ring
\cite{DHV03c,DHV03j}. These
solutions are based on circulation of
 $\elle$ tokens, where each token corresponds to a resource unit.

\vspace*{-10pt}

\paragraph{Outline.} The remainder of the paper is organized as follows: In
the next section, we define the model used in this paper.  In Section
\ref{sect:algo}, we present our self-stabilizing \klexclu\ protocol.
In Section \ref{sect:proof},
we provide the proof of correctness of our protocol,
and we analyze its waiting time.
Finally, we conclude in Section \ref{sect:ccl}.

%% file: model.tex
\section{Preliminaries}\label{sect:model}

\paragraph{Distributed Systems.} We consider {\em asynchronous distributed
systems} having a \emph{finite} number of \emph{processes}.  By asynchronous,
we mean that there is no bound on message delay, clock drift, or process
execution rate.  Every process can directly communicate with a subset of
processes called \emph{neighbors}. We denote by $\Delta_p$ the number of
neighbors of a process $p$. We consider the message-passing model where
communication between neighboring processes is carried out by
\emph{messages} exchanged through \emph{bidirectional links}, {\em i.e.},
each link can be seen as two channels in opposite directions. The
neighbor relation defines a \emph{network}. We assume that the topology of
the network is that of an \emph{oriented tree}.
\emph{Oriented} means that there is a
distinguished process called \emph{root} (denoted {\Root})
and that every non-root process knows which neighbor
is its \emph{parent} in the tree, {\em
i.e.}, the neighbor that is nearest to the root.
We say that process $q$ is a
\emph{child} of process $p$ if and only if $p$ is the parent of $q$.

A process is a sequential deterministic machine
with input\slash output capabilities and bounded local memory,
and that uses a local algorithm.
Each process executes its local algorithm by taking {\em steps}.
In a step, a process executes two actions in sequence:
(1) either it tries to receive a message
from another process, sends a message to another process, or does nothing;
and then (2) modifies some of its variables.
\footnote{When there is ambiguity, 
we denote by $x_p$ the variable $x$ in the code of process $p$.}
The local algorithm is structured as
infinite loop that contains finitely many actions.

We assume that the channels incident to a process $p$ are
locally distinguished by
a {\em label}, a number in the range $\{0 \ldots \Delta_p-1\}$;
by an abuse of notation, we may refer to a neighbor $q$ of $p$ by the label
of $p$'s channel to $q$.
We assume that the channels are {\em reliable},
meaning that no message can be lost (after transient faults are corrected)
and {\em FIFO}, meaning that messages are received in the
order they are sent.
We also assume that each channels initially contains 
some arbitrary messages, but not more than a given bound
$\CMAX$.\footnote{This assumption is required to obtain a deterministic
self-stabilizing solution working with bounded process memory;
 see \cite{GM91j}.}

A message is of the following form: $\langle\type,\valu\rangle$. The
$\valu$ field is omitted if the message does not carry any value.
A message may also contain more than one value.

A \emph{distributed protocol} is a collection of $n$ local algorithms, one
for each process. We define the {\em state} of each process to be the state of
its local memory and the contents of its incoming channels.
 The global state
of the system, referred to as a {\em configuration}, is defined as the product
of the states of processes.  We
denote by $\mathcal C$ the set of all possible configuration. An
\emph{execution} of a protocol $\mathcal P$ in a system $\mathcal
S$ is an infinite sequence of configurations (of $\mathcal S$)
$\gamma_0\gamma_1\dots\gamma_i\dots$ such that in any transition $\gamma_i
\mapsto \gamma_{i+1}$ either a process take a step, or an external ({\em
w.r.t.}
the protocol) application modifies an input variable. Any
execution is assumed to be \emph{asynchronous} but \emph{fair}: Every process
takes an infinite number of steps in the execution but the time between two
steps of a process is unbounded.

\paragraph{\klexclu.} In the {\klexclu} problem, the existence of $\elle$ units
of a shared resource is assumed.
Any process can request at most $\K$ units of
the shared resource, where $\K \leq \elle$.
We say that a protocol satisfies the \klexclu\ specification
if it satisfies the following three properties:

\BEGLIST
\item[-] {\bf Safety:} At any given time,
each resource unit ({\em n.b.},
here a resource unit corresponds to a token)
is used by at most one process,
each process uses at most $\K$ resource units, and at most $\elle$
resource units are used.
\item[-] {\bf Fairness:} If a process requests at most $\K$ resource units,
then its request is eventually satisfied
({\em i.e.} it can eventually use the resource unit it requests using a
special section of code called \emph{critical section}).
\item[-] {\bf Efficiency:}
Informally, this means that as many requests as possible must be
satisfied simultaneously.
\ENDLIST

\noindent The above mentioned notion of \emph{efficiency} is
difficult to define precisely.  A convenient parameter was
introduced in~\cite{DHV03j} to formally characterize efficiency:
{\em $(\K,\elle)$-liveness}, defined as follows.
Assume that there is a
subset $I$ of processes such that every process in $I$ is executing
its critical section forever ({\em i.e.},
it holds some resource units forever). Let $\alpha$ be the total
number of resource units held forever by the processes in $I$.
Let $R$ be the set of processes not in $I$ that are requesting
some resource units; for each $q\in R$, let $r_q$
be the number of resource units being requested by $q$, and assume that
$r_q \le \elle-\alpha$ for all $q\in R$.
Then, if $R\ne\emptyset$, at least one member of $R$
eventually satisfies its request.

\paragraph{Waiting Time.} The \emph{waiting time}~\cite{R90b} is the maximum
number of times that all processes can enter in the critical section before
some process $p$, starting from the moment $p$ requests the critical section.

\paragraph{Interface.} In any \klexclu\ protocol, a process needs to
interact with the application that requests the resource units. To manage
these interactions, we use the following interface at each process:
\BEGLIST
\item[-] $\state \in \{\request,\inside,\out\}$.
$\state = \request$\ 
means that the application is requesting some resource units. 
$\state$ switches from $\request$ to $\inside$ 
when the application is allowed
to access to the requested resource units. $\state$ switches from
$\inside$ to $\out$ when the requested resource units are released into the
system.  The switching of $\state$ from $\request$ to $\inside$
and from $\inside$ to $\out$ is managed by the \klexclu\ protocol itself;
while the switching from $\out$ to $\inside$ is managed by the application.
Other transitions (for instance, $\inside$ to $\request$) are forbidden.
\item[-] $\need \in \{0\dots \K\}$,
the number of resource units currently being requested by the application.
\item[-] $\enter$: {\em function.} This function is called by the
protocol to allow the application 
to execute the \emph{critical section}. From this call, the
application has control of the resource units 
until the end of the critical section
 (we assume that the critical section is always executed in finite, yet
unbounded, time).
\item[-] $\release$: {\em Boolean.} This predicate holds if and only if the
application is not executing its critical section.
\ENDLIST

\paragraph{Self-Stabilization~\cite{Dij74}.}
A {\em specification\/} is a predicate over the set of all executions.
A set of configurations
$\mathcal C_1 \subseteq \mathcal C$ is an {\em attractor\/} for a set of
configurations $\mathcal C_2 \subseteq \cal C$ if for any $\gamma \in C_2$
and any execution whose initial configuration is $\gamma$, the execution
contains a configuration of $\mathcal C_1$.

\begin{definition}
A protocol $\mathcal P$ is {\em self-stabilizing\/} for the specification $SP$
in a system $\mathcal S$ if there exists a non-empty subset of
$\mathcal L$ such that:
\BEGLIST
\item[-] Any execution of $\mathcal P$ in $\mathcal S$ starting from a
configuration of $\mathcal L$ satisfies $SP$ ({\em Closure Property}).
\item[-] $\mathcal L$ is an attractor for $\mathcal C$ ({\em Convergence
Property}).
\ENDLIST
\end{definition}

%% file: algo.tex
\section{Protocol}\label{sect:algo}

\input{TreeRoot}

In this section we present our self-stabilizing \klexclu\ protocol for
oriented trees (Algorithms \ref{algo3} and \ref{algo4}).
Our solution uses circulation of several types of tokens.
To clearly understand the function
of these tokens, we adopt a step-by-step approach: we start from ``naive''
non-operating circulation of $\elle$ resource tokens. Incrementally, we
augment this solution with several other types of tokens, until we obtain a
non-fault-tolerant solution.  We then add an additional control
mechanism that guarantees self-stabilization, assuming unbounded local
memory of processes.
Finally, we modify our protocol to work with bounded memory.

\paragraph{A non-fault-tolerant protocol.}

\begin{figure}[t]
\begin{center}
\epsfig{file=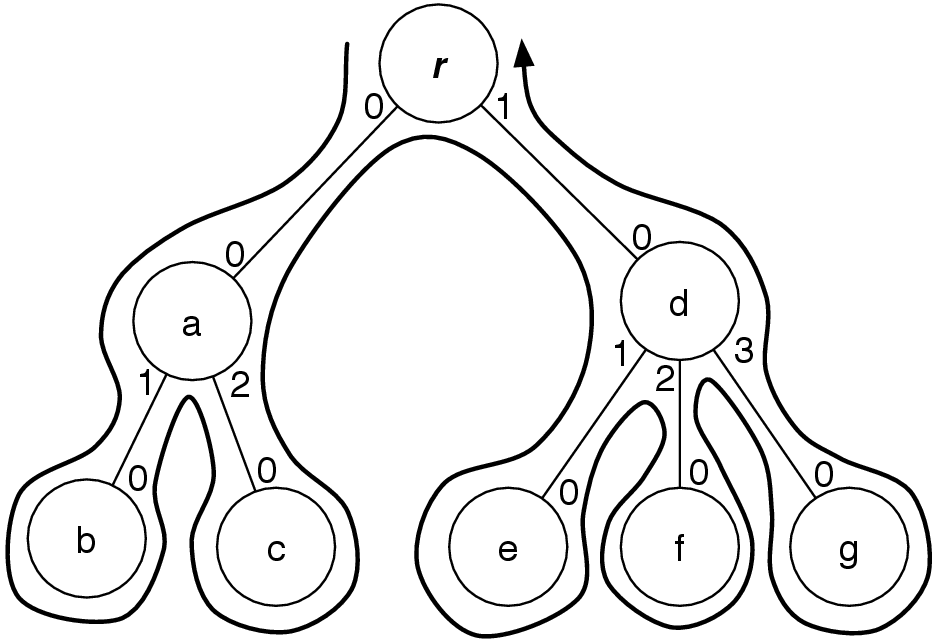, width=150pt}
\caption{\small Depth-first token circulation on oriented trees.\label{dfs}}
\end{center}
\end{figure}

\begin{figure}[htpb]
\begin{center}
\epsfig{file=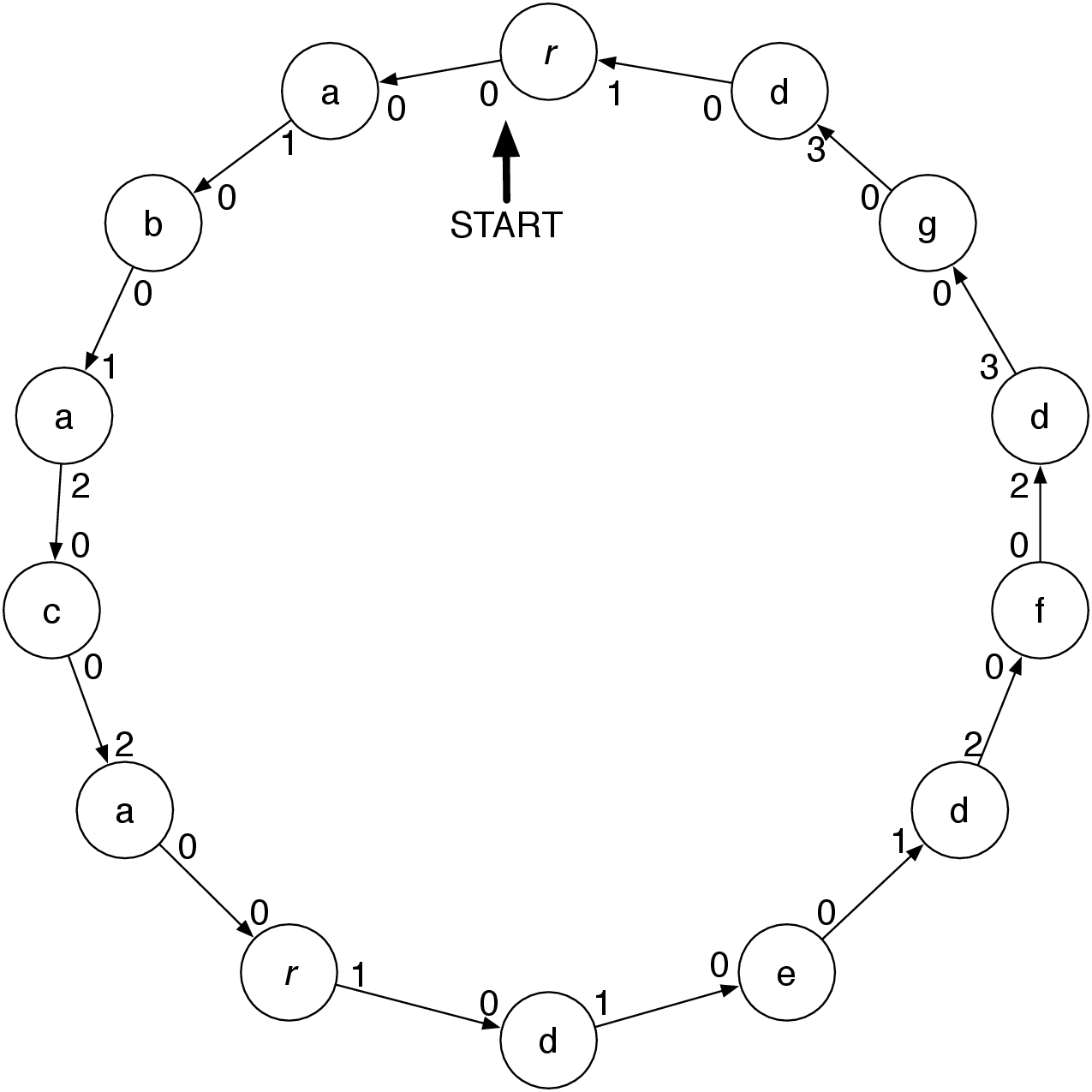, width=170pt}
\caption{\small Virtual Ring.\label{virtual}}
\end{center}
\end{figure}

The basic principle of our protocol is to use $\elle$ circulating
\emph{resource tokens} (the $\token$ messages) following depth-first
search (DFS) order: when a process $p$ receives a token from channel number
$i$, and if that token is retransmitted, either immediately or later,
it will be sent to its neighbor along channel
number $i+1$ (modulo $\Delta_p$).
(This same rule will also be followed by all
the types of tokens we will later describe.)
Figure \ref{dfs} shows the path followed by a token during
depth-first circulation in an oriented tree (recall
that any non-root process locally numbers the channel to its parent by 0).
In this way, the oriented tree emulates a ring with a designated leader
(see Figure \ref{virtual}),
and we refer to the path followed by the tokens as the {\em virtual ring\/}.

As explained Section \ref{sect:model},
the requests are managed by the variables $\state$
and $\need$. Each process also uses the multiset\footnote{{\em N.b.} a
multiset can contain several identical items.} variable $\rset$ to collect the
tokens; the collected tokens are said to be ``reserved.''
While $\state = \request$ and $|\rset| < \need$, a
process collects all tokens it receives; it also stores in $\rset$ the
number of the channel from which it receives each token,
so that when it is finally retransmitted,
it will continue its correct path around the virtual ring.
When $\state = \request$
and $|\rset| \geq \need$, it enters the critical section: $\state$ is set
to $\inside$ and the function $\enter$ is called.
Once the critical section is done
({\em i.e.}, when $\state = \inside$ and the predicate $\release$ holds)
$\state$ is set to $\out$, all tokens in $\rset$ are retransmitted,
and $\rset$ is set to $\emptyset$.
When a process receives a token it does not need,
it immediately retransmits it.

\input{TreeOther}

\begin{figure*}[htpb]
\center
\begin{multicols}{2}
\epsfig{file=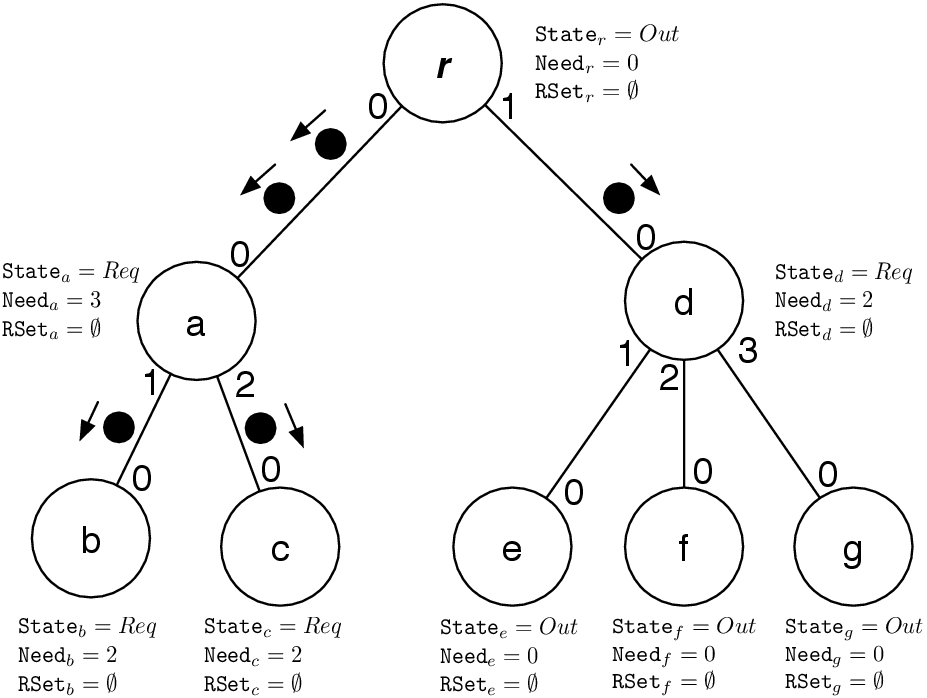, width=200pt}
\newpage
\epsfig{file=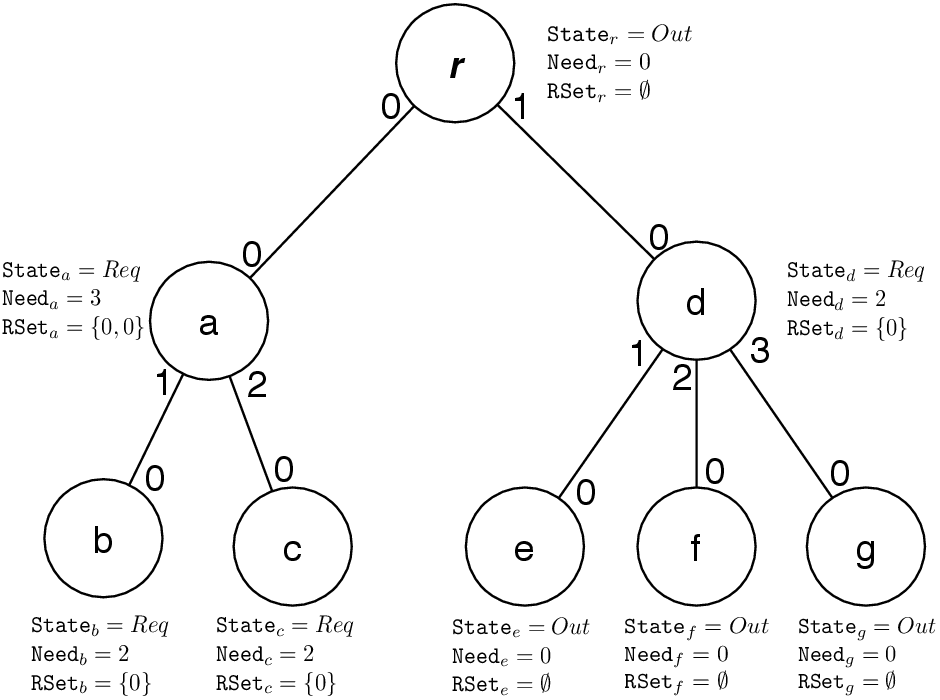, width=200pt}
\end{multicols}
\caption{\small Possible deadlock.\label{deadlock}}
\end{figure*}

Unfortunately, such a simple protocol does not always guarantee liveness.
Figure \ref{deadlock} shows a case where liveness is not maintained. In this
example, there are five resources tokens ({\em i.e.}, $\elle = 5$) and each
process can request up to three tokens ({\em i.e.}, $\K = 3$). In the
configuration shown on the left side of the figure, processes $a$, $b$,
$c$, and $d$ request more tokens than they will receive.  This
configuration will lead to the deadlock configuration shown on the right
side of the figure: processes $a$, $b$, $c$, and $d$ reserve all the tokens
they receive and never release them because their requests are never
satisfied.

We can prevent deadlock by adding a new type of token,
called the \emph{pusher} (the message $\pusher$).
If the system is in a legitimate state, there is exactly one pusher.  It
permanently circulates through the virtual ring, and prevents a process
that is not in the critical section from holding resource tokens forever.
When a process receives the pusher, it releases all its reserved tokens,
unless if it is either in its critical section ($\state = \inside$)
or is enabled to enter its critical section
($\state = \request$ and $|\rset| \geq \need$).
In either case, it retransmits the pusher.

\begin{figure*}[htpb]
\begin{center}
\epsfig{file=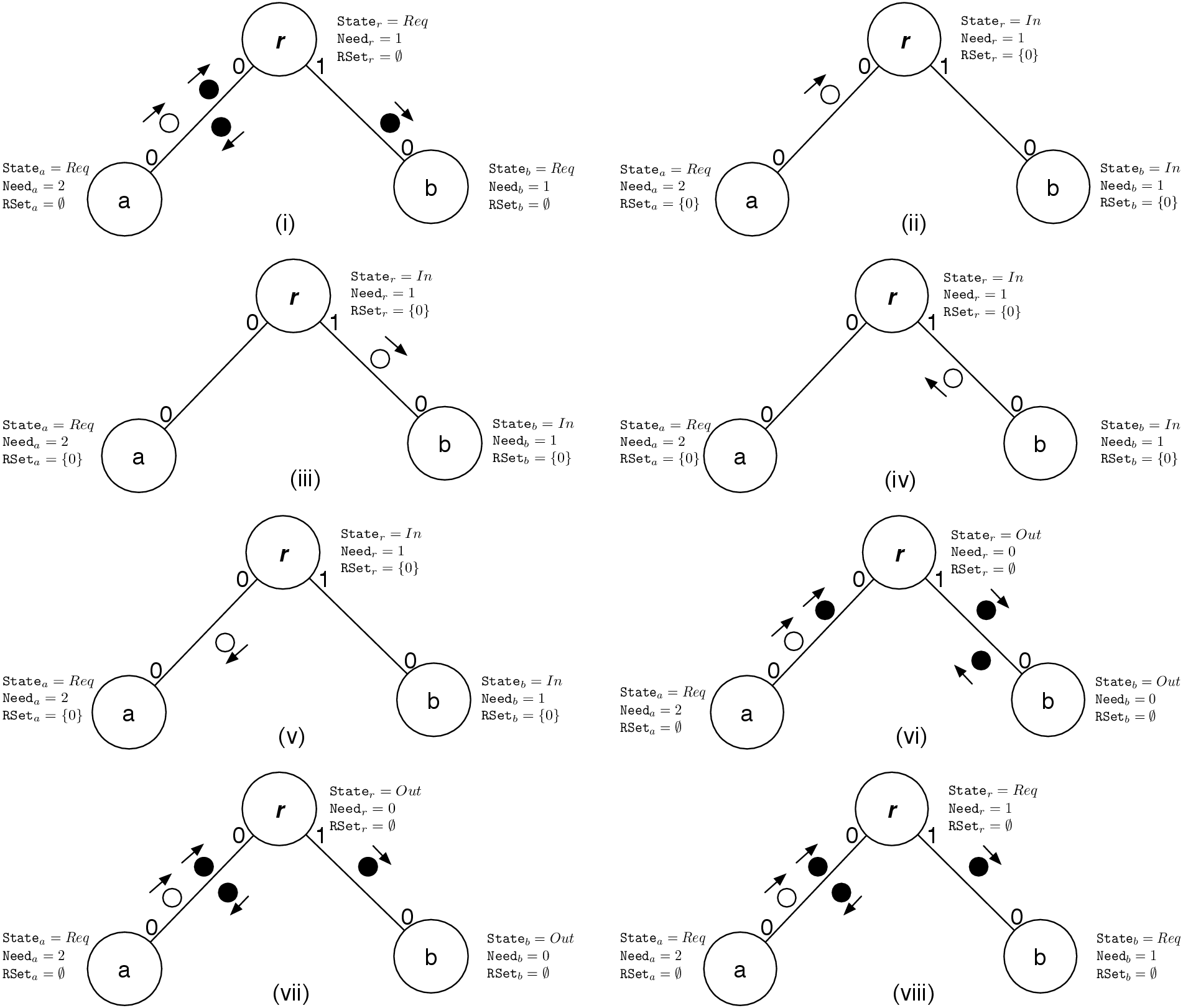, width=400pt}
\caption{\small Possible livelock.\label{livelock}}
\end{center}
\end{figure*}

The pusher protects the system from deadlock. However, it can cause
\emph{livelock}; an example is shown in Figure \ref{livelock},
for 2-out-of-3 exclusion in a tree of three processes. In Configuration $(i)$,
every process is a requester: {\Root}\ and $b$ request one resource token and
$a$ requests two resource tokens. Also, every process has a
resource token in one of its incoming channels,
and none holds any resource token.  Finally, the pusher is in the
channel from $a$ to {\Root} behind a resource token.
Every process will
collect the incoming resource token, and the system will reach the
Configuration $(ii)$ where {\Root} and $b$ execute their critical section
while $a$ is still waiting for a resource token and the pusher is reaching
{\Root}.  When {\Root} receives the pusher, it retransmits it to $b$,
while keeping its resource token, as shown in Configuration $(iii)$.
Similarly, $b$ receives the pusher while
executing its critical section, and retransmits it immediately to {\Root},
as shown in Configuration $(iv)$, after which
{\Root} retransmits the pusher to $a$ (Configuration $(v)$). Assume now
that $a$ receives the pusher while {\Root} and $b$ leave their critical
sections. We obtain Configuration $(vi)$: $a$ must release its resource
tokens because of the pusher. In Configuration $(vii)$, {\Root}
directly retransmits the resource token it receives because it is not a
requester. Finally, {\Root} and $b$ again become requesters for one resource
token in Configuration $(viii)$, which is identical to
Configuration $(i)$.  We can repeat this cycle indefinitely, and
process $a$ never satisfies its request.

To solve this problem, we add a \emph{priority token} (message $\priority$)
whose goal is to cancel the effect of the pusher.  If the system is in
a legitimate state, there is exactly one priority token.
A process which receives the priority token retransmits it immediately,
unless it has an unsatisfied request. In this case, the process holds
the priority token (the variable $\hasprio$ is set from $\perp$ to the channel
number from which the process receives the priority token)
until its request is satisfied:
the token will then be released when the process enters its critical section.
A process that holds the priority token does not release its
reserved resource tokens when it receives the pusher: it only retransmits the
pusher.  As we will show later, this guarantees
that the process will eventually satisfy its request.

Using these three types of tokens, we obtain a simple non self-stabilizing
\klexclu\ protocol.  To make it self-stabilizing,
we need additional structure.

\paragraph{A controller for self-stabilization.}
To achieve self-stabilization, we introduce one more type of token,
the {\em controller\/}.

After a finite period of transient faults,
some tokens may have disappeared or may been duplicated.
To restore correct behavior, we need an additional
self-stabilizing mechanism that regulates the number of tokens in the
network: to achieve that, we use a mechanism
similar to that introduced in~\cite{HV01c} for self-stabilizing
$\elle$-exclusion protocol on a ring. This mechanism is based on snapshot\slash reset technique.

The controller is a special token (message $\ctrl$) that
counts the other tokens; when it returns to the root after one
full circulation, the root learns the number of
tokens of each type (resource, pusher, priority), and
then adjusts these numbers as necessary.

The controller can also be effected by transient faults.
We use Varghese's \emph{counter flushing}~\cite{V00j} technique
to enforce {\em depth first token circulation\/}
(DFTC) in the tree.

We now explain how the resource tokens are counted by the controller.
(It counts the other types of tokens similarly.)
We split the count of the resource tokens into two subcounts:
\BEGLIST
\item[-] \emph{The ``passed'' tokens}. When a process holds some resource
tokens that came from channel $i$ and receives the controller from the channel
$i$, it retransmits the controller through channel $i+1$ while keeping the
resource tokens: in this case, we say that the controller {\em passes} these
tokens in the virtual ring. Indeed, these tokens were ahead the controller (in
the virtual ring) before the process received the controller, and are
behind afterward.
The field $PT$ of the controller message is used to compute the number
of the passed resource tokens.
\item[-] \emph{The tokens that are never passed by the controller}. These
tokens are counted in the variable $\stoken$ maintained at the root.
At the beginning of any circulation of the controller, the variable $\stoken$
is reset to 0. Then, until the end of the circulation of the controller, each
time a resource token starts a new circulation ({\em i.e.} the token leaves
the root from channel 0), $\stoken$ is incremented.
\ENDLIST

\noindent When the controller terminates its circulation, the number of
resource tokens in the network is equal to $PT + \stoken$,
and the numbers of pusher tokens and priority tokens is likewise known
to the root.  Three cases are then possible:
\BEGLIST
\item[-] {\em The number of tokens is correct,}
that is, there are $\ell$ resource tokens, one pusher token, and
one priority token.  In this case, the system is stabilized.
\item[-] {\em There are too few tokens.} In this case, the root creates
the number of additional tokens needed
at the end of the traversal; the system is then stabilized.
\item[-] {\em There are too many tokens of some type.}
In this case, we reset the network.
We mark the controller token with a special flag (the field $R$
in the message $\ctrl$). The root transmits the marked controller,
erases its reserved tokens as well as all the tokens it receives until
the termination of the controller's traversal.
Upon receiving the controller, every other process
erases its reserved tokens.
When the controller finishes its traversal,
there is no token in the network.
The root creates exactly $\elle$
resource tokens, one pusher token and one priority token; and we are done.
\ENDLIST

\paragraph{Self-stabilizing DFTC.}\label{ssdftc}

Using the counter flushing technique, we design a self-stabilizing $DFTC$ to
implement the controller. The principle of counter flushing is the
following: after transient faults, the token message can be lost. Hence, the
root must use a timeout mechanism to retransmit the token in case of deadlock.
The timeout is managed using the function $\restarttimer$ (that allows it to
reinitialize the timeout) and the predicate $\timeout$ (which holds
when a specified time interval is
exceeded).\footnote{We assume that this time interval
is sufficiently large to prevent congestion.}

Due to the use of the timeout, we must now deal with duplicated messages.
Furthermore,
arbitrary messages may exist in the network after faults (however they
are assumed to be bounded).
To distinguish the duplicates from the valid controller
and to flush the system of corrupted messages,
every process maintains a
counter variable $\myC$ that takes values in $\{0\dots 2(n-1)(\CMAX+1)\}$,
and marks each message with that value.
Every process also maintains a pointer $\next$ to indicate
to which process it must send the token.
The effects of the reception of a
token message differs for the root and the other processes:
\BEGLIST
\item[-] The root considers a token message as valid when the message
comes from $\next$ and
is marked with a value $c$ such that $\myC = c$. Otherwise,
it simply ignores the message, meaning it does not retransmit it.
If it receives a valid message, the root increments
$\next$ (modulo $\Delta_{\Root}$) and retransmits the token with the flag
value $\myC$ to $\next$ so that the valid token follows
DFS order. If $\next = 0$, this means that the token just finished its
previous circulation. As a consequence, the root increments $\myC$ (modulo
$2(n-1)(\CMAX+1)$) before retransmitting the token.
\item[-] A non-root process $p$ considers a message as valid in two cases: (1)
When it receives a token message from its parent (channel 0) marked with a
value $c$ such that $\myC \neq c$ or (2) when it receives a token message from
$\next$ and the message is marked with a value $c$ such that $\myC = c$. In
case (1), $p$ sets $\myC$ to $c$ and $\next$ to $\min(1,\Delta_p-1)$
({\em n.b.} in case of a leaf process $\next$ is set to 0)
before retransmitting the token message marked with $\myC$ to $\next$.
In case (2), $p$ increments $\next$ (modulo $\Delta_p$) and
then sends the token marked with $\myC$ to $\next$ so that the valid token
follows DFS order. In all other cases, $p$ considers the message to be
invalid. In the case of an invalid message coming from channel 0 with
$\myC = c$, $p$ does not consider the message in the computation, but
retransmits it to prevent deadlock.
In all other cases, $p$ simply ignores the message.
\ENDLIST
Using this method, the root increments its counter $\myC$
infinitely often and,
due to the size of the $\myC$'s domain, the $\myC$ variable of the root
eventually takes a value that does not exist anywhere else in
the system (because the number of possible values initially in the system is
bounded by $2(n-1)(\CMAX+1)$). In this case, the token marked with the new
value will be considered to be a valid token by every process.
Until the end of that traversal, the root will ignore all
other token messages.
At the end of the traversal, the system will be stabilized.

\paragraph{Dealing with bounded memory.} Due to the use of reset, the
root does not need to know the exact number of tokens at the end of the
controller's traversals. Actually, the root must only know if the number of
tokens is too high, or the number of tokens it needs to add if
the number is too low.  Hence, the
counting variables can be bounded by $\elle + 1$ for the resource tokens and
by 2 for the other types of token. The fact that a variable is assigned to its
maximum value will mean that there are too many tokens in the network
and so a reset must be started.
Otherwise, the value of the counting variable will
state whether there is a deficient number of tokens,
and in that case, how many must be added.
For any assignment to one of these bounded variables,
the value is set to the minimum between its new computed value
and the maximum value of its domain.

%% file: TreeRoot.tex
\begin{algorithm*}
\caption{\small \klexclu\ on oriented trees, code for the root \Root\label{algo3}}
\scriptsize
\begin{multicols}{2}
\begin{algorithmic}[1]
\State {\bf variables:} 
\State\qquad $C, \myC \in [0\dots 2(n-1)(\CMAX+1)]$; $\next \in [0\dots\Delta_{\Root}-1]$ 
\State\qquad $\rset$: multiset of at most $\K$ values taken in $[0\dots \Delta_{\Root}-1]$
\State\qquad $\need \in [0\dots\K]$; $\state \in \{\request$,$\inside$,$\out\}$
\State\qquad $\hasprio \in \{\perp,0,\dots,\Delta_{\Root}-1\}$
\State\qquad $R, \reset$: Booleans; $\stoken$, $PT$$\in$$[0\dots\elle+1]$
\State\qquad $\spusher, \sprio, PPr \in [0\dots$$2]$
\RepeatForever

\ForAll{$q \in [0\dots \Delta_{\Root}-1]$}

\If{(\receive{\token}{q}) $\wedge \neg \reset$}\label{line:rRT}
\If{$(\state = \request) \wedge (|\rset| < \need)$}
\State $\rset \gets \rset \cup \{q\}$
\Else
\If{$q = \Delta_{\Root}-1$}
\State $\stoken \gets \min(\stoken + 1,\elle+1)$\label{line:countRTr}
\EndIf
\State \send{\token}{q+1}
\EndIf
\EndIf\label{line:rRT2}

\If{(\receive{\pusher}{q}) $\wedge \neg \reset$}\label{line:rPT}
\If{$(\hasprio \neq \perp) \wedge (\state \neq \request \vee |\rset| < \need) \wedge$\newline
\hspace*{0.5in}$(\state \neq \inside)$}
\ForAll{$i \in \rset$}
\If{$i = \Delta_{\Root}-1$}
\State $\stoken \gets \min(\stoken + 1,\elle+1)$
\EndIf 
\State \send{\token}{i+1}
\EndFor
\State $\rset \gets \emptyset$
\EndIf
\If{$q = \Delta_{\Root}-1$}
\State $\spusher \gets \min(\spusher + 1,2)$
\EndIf
\State \send{\pusher}{q+1}
\EndIf

\If{(\receive{\priority}{q}) $\wedge \neg \reset$}\label{line:rPrT}
\If{$\hasprio = \perp$}
\State $\hasprio \gets q$
\Else
\State \send{\priority}{q+1}
\EndIf
\EndIf

\If{(\receive{\ctrl,C,R,PT,PPr}{q})}

\If{$(q = \next) \wedge (\myC = C)$}

\State $\next \gets \next + 1$

\If{$\next = 0$}
\State $\myC \gets \myC + 1$
\State $\reset \gets (PT + \stoken > \elle) \vee$\newline
\hspace*{0.8in} $(PPr + \sprio > 1) \vee (\spusher > 1)$\label{line:set:reset}

\If{$\reset$}\label{line:resetR}
\State $\rset \gets \emptyset$\label{line:reset:rsetR}
\State $\hasprio \gets \perp$
\Else

\If{$PPr + \sprio < 1$}
\State \send{\priority}{0}
\EndIf

\While{$PT + \stoken < \elle$}\label{line:complement:deb}
\State \send{\token}{0}
\State $\stoken \gets \min(\stoken + 1,\elle+1)$
\EndWhile \label{line:complement:fin}

\If{$\spusher < 1$}
\State \send{\pusher}{0}
\EndIf

\EndIf

\State $\stoken \gets 0$\label{line:reinit:stoken}
\State $\sprio \gets 0$
\State $\spusher \gets 0$
\State $PT \gets 0$\label{line:reinit:PT}
\State $PPr \gets 0$

\EndIf
\State $PT \gets \min(PT + |\rset|_{q},\elle+1)$\label{line:set:PTR}
\If{$\hasprio = q$}
\State $PPr \gets \min(PPr + 1,2)$
\EndIf
\State \send{\ctrl,\myC,\reset,PT,PPr}{\next}\label{line:newCT}
\State $\restarttimer$
\EndIf
\EndIf

\EndFor

\If{$(\state = \request) \wedge (|\rset| \geq \need)$}
\State $\state \gets \inside$
\State $\enter$
\EndIf

\If{$(\state = \inside) \wedge \release$}
\ForAll{$i \in \rset$}
\If{$i = \Delta_{\Root}-1$}
\State $\stoken \gets \min(\stoken + 1,\elle+1)$
\EndIf 
\State \send{\token}{i+1}
\EndFor
\State $\rset \gets \emptyset$
\State $\state \gets \out$
\EndIf

\If{$(\hasprio \neq \perp) \wedge (\state \neq \request \vee |\rset| \geq \need)$} \label{rsendprio}
\If{$\hasprio = \Delta_{\Root}-1$}
\State $\sprio \gets \min(\sprio + 1,2)$
\EndIf
\State \send{\priority}{\hasprio+1}
\State $\hasprio \gets \perp$
\EndIf

\If{$\timeout$}
\State \send{\ctrl,\myC,\reset,0,0}{\next}
\State $\restarttimer$
\EndIf

\EndRepeatForever
\end{algorithmic}
\end{multicols}
\end{algorithm*}

%% file: TreeOther.tex
\begin{algorithm*}[t]
\caption{\small \klexclu\ on oriented trees, code for the other process $p$\label{algo4}}
\scriptsize
\begin{multicols}{2}
\begin{algorithmic}[1]
\State {\bf variables:} 
\State\qquad $C,\myC \in [0\dots 2(n-1)(\CMAX+1)]$; $\next \in [0\dots\Delta_p-1]$ 
\State\qquad $\rset$: multiset of at most $\K$ values taken in $[0\dots \Delta_p-1]$
\State\qquad $\need \in [0\dots\K]$; $\state \in \{\request$,$\inside$,$\out\}$
\State\qquad $\hasprio \in \{\perp,0,\dots,\Delta_p - 1\}$
\State\qquad $R,\ok$: Booleans; $PT$$\in$$[0\dots\elle+1]$; $PPr$$\in$$[0\dots 2]$
\RepeatForever

\ForAll{$q \in [0\dots \Delta_p-1]$}
\If{(\receive{\token}{q})}\label{line:pRT}
\If{$(\state = \request) \wedge (|\rset| < \need)$}
\State $\rset \gets \rset \cup \{q\}$ 
\Else 
\State \send{\token}{q+1}
\EndIf
\EndIf\label{line:pRT2}

\If{(\receive{\pusher}{q})}
\If{$(\hasprio \neq \perp) \wedge (\state \neq \request \vee |\rset| < \need) \wedge$\newline
\hspace*{0.5in} $(\state \neq \inside)$}
\ForAll{$i \in \rset$} 
\State \send{\token}{i+1}
\EndFor
\State $\rset \gets \emptyset$
\EndIf
\State \send{\pusher}{q+1}
\EndIf

\If{(\receive{\priority}{q})}
\If{$\hasprio = \perp$}
\State $\hasprio \gets q$
\Else
\State \send{\priority}{q+1}
\EndIf
\EndIf

\If{(\receive{\ctrl,C,R,PT,PPr}{q})}
\State $\ok \gets false$
\If{$(q = \next) \wedge (\myC = C) \wedge (\next \neq 0)$}
\State $\next \gets \next + 1$
\State $\ok \gets true$
\If{$R$}\label{line:resetP1}
\State $\rset \gets \emptyset$\label{line:reset:rsetP1}
\State $\hasprio \gets \perp$
\EndIf
\EndIf
\If{$(q = 0)$}
\State $\ok \gets true$
\If{$\myC \neq C$}
\State $\next \gets \min(1,\Delta_p-1)$ 
\If{$R$}\label{line:resetP2}
\State $\rset \gets \emptyset$\label{line:reset:rsetP2}
\State $\hasprio \gets \perp$
\EndIf
\EndIf
\State $\myC \gets C$
\EndIf
\If{$\ok$}
\State $PT \gets \min(PT+|\rset|_{q},\elle+1)$\label{line:set:PTP}
\If{$\hasprio = q$}
\State $PPr \gets \min(PPr + 1,2)$
\EndIf
\State \send{\ctrl,\myC,R,PT,PPr}{\next}
\EndIf
\EndIf

\EndFor

\If{$(\state = \request) \wedge (|\rset| \geq \need)$}
\State $\state \gets \inside$
\State $\enter$
\EndIf

\If{$(\state = \inside) \wedge \release$}
\ForAll{$i \in \rset$} 
\State \send{\token}{i+1}
\EndFor
\State $\rset \gets \emptyset$
\State $\state \gets \out$
\EndIf

\If{$(\hasprio \neq \perp) \wedge (\state \neq \request \vee |\rset| \geq \need)$} \label{sendprio}
\State \send{\priority}{\hasprio+1}
\State $\hasprio \gets \perp$
\EndIf

\EndRepeatForever
\end{algorithmic}
\end{multicols}
\end{algorithm*}

%% file: preuve.tex
\section{Correctness and Waiting Time}\label{sect:proof}

In this section, we first prove that our protocol is a self-stabilizing
\klexclu\ protocol. We then analyze its waiting time.

\paragraph{Correctness.}

We split the proof into three steps. (1) Recall that the controller part of
our protocol is a self-stabilizing DFS token circulation. (2) We show that
once the controller is stabilized to DFS token circulation, the system
eventually stabilizes to the expected number of different tokens. (3) We show
that once the system contains the expected number of tokens,
the system stabilizes to the \klexclu\ specification.

In our protocol, when a process receives a $\ctrl$ message, either it
considers the message as valid or not. The process takes account of the
messages for computations only when they are valid. Assume that a
process $p$ receives a $\ctrl$ message marked with the flag value $c$ from
channel $q$. Process $p$ considers this message as valid if and only if
$(q=\next_p \wedge c=\myC_p) \vee (p \neq \Root \wedge (q=0 \wedge
c\neq\myC_p))$.

In the following, we call any $\ctrl$ message a {\em control token}.
Each time a
process receives a valid $\ctrl$ message, it makes some local computations, and
then sends another $\ctrl$ message. In the case of a non-root process, the
sent message is marked with the same flag as the received message: we consider
it to be the same control token. In
the case of the root, the sent message is marked either with the same value or
with a new one. In the former case, we consider it to be the same
control token, while in the latter case, we consider the received
control token to have terminated its traversal, and the transmitted
control token to be new.

To implement the control part, we use the counter flushing techniques
introduced by Varghese in \cite{V00j}. Hence, from \cite{V00j}, we can deduce
the following lemma:

\begin{lemma}\label{lem:dfstc}
Starting from any configuration, the system
converges to a configuration at which:
\BEGLIST
\item[1.] There exists at most one valid control token in the network.
\item[2.] The root regularly creates a new valid control token.
\item[3.] Any valid control token visits all processes in DFS order.
\ENDLIST
\end{lemma}

\begin{remark}
In our protocol, only the valid control tokens are considered in the
computations. Hence, from now on, we only consider the valid control
tokens and we simply refer to them as {\em control tokens}.
\end{remark}

\noindent We now show that starting from any configuration, the system
eventually contains the expected number of each type of token.

Note that each resource token is either in a link (in this case, the token is
said to be {\em free}) or it is stored in the $\rset$ of a process (in this
case, the token is said to be {\em reserved}). Hence, at any time the number
of resource tokens in the network is equal to the sum of the sizes of the
$\rset$ multisets plus the number of {\em free} resource tokens.

Similarly, at any time, the number of priority tokens is equal to the
number of processes satisfying $\hasprio \neq \perp$ plus the number of {\em
free} priority tokens.

Finally, as a process cannot store any pusher token, the number of pusher
tokens is equal to the number of free pusher tokens.

\begin{lemma}\label{lem:reset}
Let $\gamma$ be the first configuration after the control part is stabilized.
If, after $\gamma$, the root creates a control token whose reset field $R$
is true, then the system contains no resource, priority, and pusher
token at the end of the traversal of the control token.
\end{lemma}
\begin{proof}
Consider any control token created by the root after configuration $\gamma$.
Assume that the reset field $R$ of the control token is set to true. Then,
the $\reset$ variable of the root is also true (see Line \ref{line:newCT} in
Algorithm \ref{algo3}). $\reset_\Root$ remains true until the control token
terminates its traversal. Hence, during the traversal, any token (except the
control token) that is received by
the root is ignored by the root and so disappears from the network (see Lines
\ref{line:rRT}, \ref{line:rPT}, and \ref{line:rPrT} in Algorithm \ref{algo3}).
Also, during its traversal, each process erases all
tokens (except the control token) it holds when visited by the control
token (see Line \ref{line:resetR} in Algorithm \ref{algo3}, and Lines
\ref{line:resetP1}, and \ref{line:resetP2} in Algorithm \ref{algo4}). Hence,
every resource, priority, or pusher token is either erased at a process when
the process is visited by the control token, or is pushed to the root and
then disappears.  At the end of the traversal of the control token, the system
contains no resource, priority, or pusher tokens.
\end{proof}

\begin{lemma}\label{lem:count:R}
Let $\gamma$ be the first configuration after the control part is stabilized.
When a control token created by the root after $\gamma$ terminates its
traversal, we have:
\BEGLIST
\item[-] If $PT + \stoken_\Root > \elle$,
then there are more than $\elle$ resource tokens in the network.
\item[-] If $PT + \stoken_\Root \leq \elle$,
then there are exactly $PT + \stoken$ resource tokens in the network.
\ENDLIST
\end{lemma}
\begin{proof}Consider any control token created by the root after
configuration $\gamma$. There are two cases:
\BEGLIST
\item[-] {\em The reset field $R$ of the control token is true.} By Lemma
\ref{lem:reset}, there is no resource token in the network when the control
token terminates its circulation. So, to prove the lemma in this case, we
must show that $PT + \stoken_\Root = 0$ at the end of the circulation.

First, $\stoken_\Root$ is reset to 0 (Line \ref{line:reinit:stoken}) before the
control token starts its circulation (Line \ref{line:newCT}). Also, $\reset_\Root$
is true when the control token starts its circulation (see Line
\ref{line:newCT} in Algorithm \ref{algo3}). Thus, until termination of the
circulation, {\Root} ignores any resource tokens it receives (see Lines
\ref{line:rRT}, \ref{line:rPT}, and \ref{line:rPrT}) and so $\stoken_\Root$ is
still equal to 0 at the end of the control token circulation.

Consider now the $PT$ field of the control token. Before
the start of the control token circulation, {\Root} executes the following
action: $\rset$ is set to $\emptyset$ (Line \ref{line:reset:rsetR} in
Algorithm \ref{algo3}), $PT$ is set to 0 (Line \ref{line:reinit:PT} in
Algorithm \ref{algo3}), and, as a consequence, $PT$ is set to
$\min(0,\elle+1)$ (Line \ref{line:set:PTR} in Algorithm \ref{algo3}).
So, at the start of the control token circulation, 
the control token is sent with its $PT$ field equal to 0.
Since the reset field $R$ of the control token is true, each
time the control token arrives at a process,
the process resets its $\rset$ variable to $\emptyset$ (see Lines
\ref{line:reset:rsetR} in Algorithm \ref{algo3}, Lines
\ref{line:reset:rsetP1}, and \ref{line:reset:rsetP2} in Algorithm \ref{algo4})
before setting $PT$ to $\min(PT+|\rset|_q,\elle+1)$ (see Line
\ref{line:set:PTR} in Algorithm \ref{algo3} and Line \ref{line:set:PTP} in
Algorithm \ref{algo4}) and then retransmitting the token. Hence, $PT$ remains
equal to 0 until the end of the circulation.

When the control token terminates its circulation, $PT +
\stoken_\Root = 0$, and we are done.
\item[-] {\em The reset field $R$ of the control token is false.}
In this
case, we can remark that no resource token is erased during the circulation of
the control token, because both $R$ and $\reset_\Root$ are false.

(*) We now show that any resource token is counted at most once during the
circulation of the control token.
Due to the FIFO quality of the links and the fact that when the
control token is received by a process, the process receives no other message
before retransmitting the control token, we have
the following property: a resource token is passed by the control token at
most once during a circulation. So, during the circulation, either the
resource token is counted into the $PT$ field of the control token when the
resource token is passed by the control token (see Line \ref{line:set:PTR} in
Algorithm \ref{algo3} and Line \ref{line:set:PTP} in Algorithm \ref{algo4}) or
it is counted at the root when it terminates a loop of the virtual ring (Line
\ref{line:countRTr}). Hence, any resource token is counted at most once.

(**) Finally we show, by contradiction, that any resource token is counted
at least once during the circulation of the control token.
Assume that a resource token is not counted during that circulation.
Then, the resource token is never passed by the control token.
The links are FIFO, and when the control token is received by a process, the
process receives no other message before retransmitting
the control token. So, the resource token is always ahead the control token in
the virtual ring. As a consequence, the resource token is eventually counted
at the root when it terminates a loop of the virtual ring (Line
\ref{line:countRTr}), contradiction.

From (*), we know that if $PT + \stoken_\Root > \elle$ at the end of the
control token circulation, then there are more that $\elle$ resource tokens
in the network. From (*) and (**), we know that if $PT + \stoken_\Root \leq
\elle$ at
the end of the control token circulation, then there are exactly $PT +
\stoken_\Root$ resource tokens in the network, and we are done.
\ENDLIST
\end{proof}

\noindent Following similar reasoning, we obtain the following two lemmas:

\begin{lemma}\label{lem:count:Pr}
Let $\gamma$ be the first configuration after the control part is stabilized.
When a control token created by the root after $\gamma$ terminates its
traversal, we have:
\BEGLIST
\item[-] If $\sprio_\Root + PPr > 1$, there is more that one priority token
in the network.
\item[-] If $\sprio_\Root + PPr \leq 1$, there are exactly $\sprio + PPr$
priority tokens in the network.
\ENDLIST
\end{lemma}

\begin{lemma}\label{lem:count:P}
Let $\gamma$ be the first configuration after the control part is stabilized.
When a control token created by the root after $\gamma$ terminates its
traversal, we have:
\BEGLIST
\item[-] If $\spusher_\Root > 1$, then there is more that one pusher token in
the network.
\item[-] If $\spusher_\Root \leq 1$, then there
are exactly $\spusher_\Root$ pusher tokens
in the network.
\ENDLIST
\end{lemma}

\begin{lemma}\label{lem:tokenR}
Starting from any configuration, the system eventually reaches a configuration
from which there always exist exactly $\elle$ resources tokens.
\end{lemma}
\begin{proof}
Let $\gamma$ be the first configuration after the control part is stabilized.
Consider any control token created by the root after $\gamma$. Let us study
the two following cases:
\BEGLIST
\item[-] {\em $PT + \stoken_\Root \leq \elle$ at the end of the control token
traversal.} Then, $\reset_\Root$ is set to false.
(Line \ref{line:set:reset} in
Algorithm \ref{algo3}) and, as a consequence, the reset field of the next
control token will be false
(Line \ref{line:newCT} of Algorithm
\ref{algo3}). Hence, no resource token will be erased during
the next circulation of a control token. If $PT + \stoken_\Root < \elle$, then
exactly $\elle - (PT + \stoken_\Root)$ are created (see Lines
\ref{line:complement:deb} to \ref{line:complement:fin} in Algorithm
\ref{algo3}). Hence, the number of resource tokens will be exactly equal to
$\elle$ at the beginning of
the next control token circulation. By Lemma \ref{lem:count:R}, $PT +
\stoken_\Root$ will be equal to $\elle$ at the end of the next control token
circulation, no resource token will be added.  Any later circulation
of the control token cannot change the number of resource tokens.
Hence, the system will contain $\elle$ resource tokens forever.

\item[-] {\em $PT + \stoken_\Root > \elle$ at
the end of the control token traversal.} Then, $\reset_\Root$ is set to true
(Line \ref{line:set:reset} in Algorithm \ref{algo3}) and, as a consequence,
the reset field of the next control token will be true (Line
\ref{line:newCT} of Algorithm \ref{algo3}). By Lemmas \ref{lem:reset} and
\ref{lem:count:R}, reducing to the previous case when
the circulation of the next control token terminates, and we are done.
\ENDLIST
\end{proof}

\noindent Following similar reasoning, we can deduce from Lemmas
\ref{lem:reset}, \ref{lem:count:Pr}, and \ref{lem:count:P}, the following two 
lemmas:

\begin{lemma}\label{lem:tokenPr}
Starting from any configuration, the system eventually reaches a configuration
from which there always exists one priority token.
\end{lemma}

\begin{lemma}\label{lem:tokenP}
Starting from any configuration, the system eventually reaches a configuration
after which there always exists one pusher token.
\end{lemma}

\noindent We now show that once the system contains the correct number
of each type of token, the system stabilizes to the \klexclu\ specification.

\begin{lemma}\label{infty:P}
Starting from any configuration, every process receives a pusher token
infinitely many times.
\end{lemma}
\begin{proof}By Lemmas \ref{lem:tokenR}, \ref{lem:tokenPr}, and
\ref{lem:tokenP}, starting from
any configuration, the system eventually reaches a
configuration $\gamma$ after which there are always
 exactly $\elle$ resource tokens,
one priority token, and one pusher token in the network. From $\gamma$, the
system is then never again reset.  So, from $\gamma$,
the unique pusher token of the system always follows DFS order.
Each time a process receives the pusher token, it retransmits it in finite
time.  Hence, every processes receives it infinitely often and the lemma holds.
\end{proof}

\begin{lemma}\label{infty:R}
Starting from any configuration, every process receives a priority token
infinitely many times.
\end{lemma}
\begin{proof}
By Lemmas \ref{lem:tokenR}, \ref{lem:tokenPr}, and \ref{lem:tokenP},
 starting from
any configuration, the system eventually reaches a configuration $\gamma$ from
which there are $\elle$ resource tokens, one priority token, and one pusher
token in the network. From $\gamma$, the system is never again reset.
So from $\gamma$, the unique priority token of the system always follow the
DFS order.

By way of contradiction, assume that, from $\gamma$, a process eventually
stops receiving the priority token. Since the priority token
circulates in DFS order and traverses each link in finite time,
we can deduce that some other process $p$ eventually holds it forever. In this
case, $p$ is a requester and its request is never satisfied. Now, by Lemma
\ref{infty:P} other process receives
the pusher token infinitely often. So, each other process retransmits
the resource tokens it holds within finite time, because it eventually
either satisfies its request, executes its critical sections,
and then releases its tokens, or does not satisfy its request, but receives
the pusher, and then releases its resource tokens.
Similarly the resource tokens always follows DFS
order after $\gamma$.
Hence, $p$ receives resource tokens infinitely many times,
and never releases the the priority token even if it receives
the pusher token.
Since $\K \leq \elle$, the request of $p$ is eventually
satisfied, contradiction.
\end{proof}

\begin{lemma}\label{infty:Pr}
Starting from any configuration, every process receives resource tokens
infinitely many times.
\end{lemma}
\begin{proof}
By Lemmas \ref{lem:tokenR}, \ref{lem:tokenPr}, and \ref{lem:tokenP},
 starting from
any configuration, the system eventually reaches a configuration $\gamma$ from
which there are $\elle$ resource tokens, one priority token, and one pusher
token in the network.
After $\gamma$, the system is then never again reset. Thus, after
$\gamma$, the resource tokens of the system always follow DFS order.

Assume, by way of contradiction, that some process only receives
resource tokens finitely many times.
This implies that every resource token is eventually held forever by some
process.
Consider one process that holds at least one resource token forever.
By Lemma \ref{infty:P},
that process cannot hold the priority token forever.
 When it releases the priority token, either
its request is satisfied, it executes the critical section within finite
time, and then releases its resource tokens, or it is not a requester
and thus must release its resource token.  Either case is a contradiction,
and we are done.
\end{proof}

\begin{lemma}\label{lem:spec:2}Starting from any configuration, the fairness
property of the \klexclu\ specification is eventually satisfied.
\end{lemma}
\begin{proof}Assume that there a request by some process $p$ that is never
satisfied.
By Lemmas \ref{lem:tokenR}, \ref{lem:tokenPr}, and \ref{lem:tokenP}, starting
from
any configuration, the system eventually reaches a configuration $\gamma$ from
which there are always
 $\elle$ resource tokens, one priority token, and one pusher
token into the network. After $\gamma$, the system is then never again reset.
Hence,after $\gamma$, if $p$ holds
the priority token, it releases it only if its request is satisfied.
By Lemma \ref{infty:Pr}, $p$ eventually receives the priority token.
Again by Lemma \ref{infty:Pr},
$p$ eventually releases the priority token, and so its
request must have been satisfied, contradiction.
\end{proof}

\begin{lemma}\label{lem:spec:1}Starting from any configuration, the safety
property of the \klexclu\ specification is eventually satisfied.
\end{lemma}
\begin{proof}
By Lemma \ref{lem:tokenR}, there are eventually exactly $\elle$ resource tokens
in the network. Hence, eventually, exactly $\elle$ resource unit are available
in the system.

Finally, any process $p$ that initially holds some resource tokens eventually
releases them because either is is not a requester or it eventually satisfies
its request by Lemma \ref{lem:spec:2}. Hence, eventually $p$ sets $\rset$ to
$\emptyset$ and
then $|\rset| \leq \need$ forever because each time $p$ receives a resource
token while $|\rset| \geq \need$, it directly retransmits it (see Lines
\ref{line:rRT} to \ref{line:rRT2} in Algorithm \ref{algo3} and Lines
\ref{line:pRT} to \ref{line:pRT2} in Algorithm \ref{algo4}). Now, $\need$ is
always less than or equal to $\K$. Hence, every process eventually only uses at
most $\K$ resource tokens (units) simultaneously.
\end{proof}

\begin{lemma}\label{lem:spec:3}Starting from any configuration, the efficiency
property of the \klexclu\ specification is eventually satisfied.
\end{lemma}
\begin{proof}
We use the definition of efficiency given in~\cite{DHV03j}.
We prove that starting from any configuration,
{\em $(\K,\elle)$-liveness} is eventually satisfied.

Consider the configuration $\gamma$ after which:
(1) there are always $\elle$ resource tokens, one
priority token, and one pusher token; and (2) the safety properties of the
\klexclu\ are satisfied (such a configuration exists by Lemmas
\ref{lem:tokenR}, \ref{lem:tokenPr}, \ref{lem:tokenP}, and \ref{lem:spec:1}).

Assume that after $\gamma$,
the system reaches a configuration $\gamma^\prime$
after which there is a subset $I$ of processes such that every process in $I$
executes its critical section forever (in this case they hold some resource
units forever). Let $\alpha$ be the total
number of resource units held forever by the
processes in $I$.

Assume then that there are some processes not in $I$ that request some resource
units and each of these processes requests at most $\elle - \alpha$ resource
units.

The priority token follows DFS order. Since every process in $I$
executes the critical section forever, none of these processes keeps the
priority token forever (see Lines \ref{rsendprio} in Algorithm \ref{algo3} and
\ref{sendprio} in Algorithm \ref{algo4}). Finally, every non-requester
directly retransmits the priority token when
it receives it (see Line \ref{rsendprio} in Algorithm \ref{algo3} and Line
\ref{sendprio} in Algorithm \ref{algo4}). Hence, there is a requesting process
$p$ which is not in $I$ that eventually receives the priority token. From that
point, $p$ will release it only after its request is satisfied (see Line
\ref{rsendprio} in Algorithm \ref{algo3} and Line \ref{sendprio} in Algorithm
\ref{algo4}). As a consequence, $p$ will keep every resource token it receives,
even if it receives the pusher token.
Checking the proof of Lemma \ref{infty:P}, we can
see that Lemma \ref{infty:P} still holds even if some processes execute the
critical section forever. So, by Lemma \ref{infty:P} every process that is
not in $I \cup \{p\}$ receives the pusher token infinitely often, and so
cannot hold resource tokens forever. Finally, every process in $I$
directly retransmits the resource tokens it receives
while it is executing the critical section
because they satisfy $|\rset| \geq \need$ by Lemma \ref{lem:spec:1} (see Lines
\ref{line:rRT} to \ref{line:rRT2} in Algorithm \ref{algo3} and Lines
\ref{line:pRT} to \ref{line:pRT2} in Algorithm \ref{algo4}). So, $p$
eventually receives the resource tokens it needs to perform the critical
section (remember that $p$ requests at most $\elle - \alpha$ resource units)
and we are done.
\end{proof}

\noindent From Lemmas \ref{lem:spec:1}, \ref{lem:spec:2}, and \ref{lem:spec:3},
we obtain:

\begin{theorem}
The protocol proposed in Algorithms \ref{algo3} and \ref{algo4}
is a self-stabilizing \klexclu\ protocol for tree networks.
\end{theorem}

\paragraph{Waiting Time.}

\begin{theorem}
Once the protocol proposed in Algorithms \ref{algo3} and \ref{algo4} is
stabilized, the waiting time is $\elle \times (2n-3)^2$ in the worst case.
\end{theorem}
\begin{proof}
We first show that the waiting time of a requesting process that holds the
priority token is $\elle \times (2n-3)$ in the worst case.
Consider a process
$p$ that requests some resource units and holds the priority token. In the
worst case, $p$ appears only once in the virtual ring defined by the DFS order
(if $p$ is a leaf). Also in
the worst case, the $\elle$ resource tokens may traverse the entire
virtual ring before $p$ receives
the tokens it needs.
The virtual ring can contain up to $(2n-3)$
processes in addition to $p$.
Any resource token may satisfy one request each time it traverses a
process (in the worst case,
each process other than $p$ always requests one token).
Hence, the $\elle$ resource tokens
may satisfy up to $\elle \times (2n-3)$ requests before $p$ satisfies its
request.

Using similar reasoning, we can see that
a requesting process could wait until the priority
token traverses the whole virtual ring (up to $2(n-2)$ nodes) before it
satisfy its request; during that time, up to  $\elle \times (2n-3)^2$ requests
can be satisfied, and we are done.
\end{proof}

%% file: ccl.tex
\section{Conclusion and Perspectives}\label{sect:ccl}

In this paper, we propose the first (deterministic) self-stabilizing
distributed \klexclu\  protocol for asynchronous oriented tree networks. The
proposed protocol uses a realistic model of computation, the message-passing
model. The only restriction we make is to assume that the channels initially
contain a bounded known number of arbitrary messages. We make this assumption
to obtain a solution that uses bounded memory per process
(see the results in~\cite{GM91j}). However, if we assume unbounded
process memory, our solution can be easily adapted to work without
assumptions on channels (following the method presented in~\cite{KP93j}).

The main interest in dealing with an oriented tree is that solutions
on the oriented tree can
be directly mapped to solutions for arbitrary rooted networks
by composing the protocol
with a spanning tree construction ({\em e.g},~\cite{AB97j,DDT05c}).

There are several possible extensions of our work.
On the theoretical side, one can
investigate whether the waiting time of our solution
($\elle \times (2n-3)^2$) can be improved.
Possible extension to networks where processes are subject
to other failure patterns, such as process crashes, remains open.
On the practical side, our solution is designed in a realistic model and can
be extended to arbitrary rooted networks. Hence, implementing our solution in
a real network is a future challenge.

%% file: techrep.bbl
\begin{thebibliography}{10}

\bibitem{AB97j}
Y~Afek and A~Bremler.
\newblock Self-stabilizing unidirectional network algorithms by power supply.
\newblock {\em Chicago Journal of Theoretical Computer Science}, 1998:Article
  3, 1998.

\bibitem{DHV03c}
A~K Datta, R~Hadid, and V~Villain.
\newblock A new self-stabilizing k-out-of-l exclusion algorithm on rings.
\newblock In Shing-Tsaan Huang and Ted Herman, editors, {\em Self-Stabilizing
  Systems}, volume 2704 of {\em Lecture Notes in Computer Science}, pages
  113--128. Springer, 2003.

\bibitem{DHV03j}
A~K Datta, R~Hadid, and V~Villain.
\newblock A self-stabilizing token-based k-out-of-l exclusion algorithm.
\newblock {\em Concurrency and Computation: Practice and Experience},
  15(11-12):1069--1091, 2003.

\bibitem{DDT05c}
Sylvie Dela{\"e}t, Bertrand Ducourthial, and S{\'e}bastien Tixeuil.
\newblock Self-stabilization with r-operators revisited.
\newblock In Ted Herman and S{\'e}bastien Tixeuil, editors, {\em
  Self-Stabilizing Systems}, volume 3764 of {\em Lecture Notes in Computer
  Science}, pages 68--80. Springer, 2005.

\bibitem{Dij74}
EW~Dijkstra.
\newblock Self stabilizing systems in spite of distributed control.
\newblock {\em Communications of the Association of the Computing Machinery},
  17:643--644, 1974.

\bibitem{FLBB89j}
M~J Fischer, N~A Lynch, J~E Burns, and A~Borodin.
\newblock Distributed fifo allocation of identical resources using small shared
  space.
\newblock {\em ACM Trans. Program. Lang. Syst.}, 11(1):90--114, 1989.

\bibitem{GM91j}
Mohamed~G. Gouda and Nicholas~J. Multari.
\newblock Stabilizing communication protocols.
\newblock {\em IEEE Trans. Computers}, 40(4):448--458, 1991.

\bibitem{HV01c}
Rachid Hadid and Vincent Villain.
\newblock A new efficient tool for the design of self-stabilizing l-exclusion
  algorithms: The controller.
\newblock In Ajoy~Kumar Datta and Ted Herman, editors, {\em WSS}, volume 2194
  of {\em Lecture Notes in Computer Science}, pages 136--151. Springer, 2001.

\bibitem{KP93j}
Shmuel Katz and Kenneth~J. Perry.
\newblock Self-stabilizing extensions for message-passing systems.
\newblock {\em Distributed Computing}, 7(1):17--26, 1993.

\bibitem{MBRA98j}
Y~Manabe, R~Baldoni, M~Raynal, and S~Aoyagi.
\newblock k-arbiter: A safe and general scheme for h-out of-k mutual exclusion.
\newblock {\em Theor. Comput. Sci.}, 193(1-2):97--112, 1998.

\bibitem{MT99c}
Y~Manabe and N~Tajima.
\newblock ({\it k})-arbiter for {\it h}-out of-{\it k} mutual exclusion
  problem.
\newblock In {\em ICDCS}, pages 216--223, 1999.

\bibitem{R91c}
M~Raynal.
\newblock A distributed solution to the k-out of-m resources allocation
  problem.
\newblock In F~K H~A Dehne, F~Fiala, and W~W Koczkodaj, editors, {\em ICCI},
  volume 497 of {\em Lecture Notes in Computer Science}, pages 599--609.
  Springer, 1991.

\bibitem{R86b}
Michel Raynal.
\newblock {\em Algorithms for Mutual Exclusion}.
\newblock MIT Press, 1986.

\bibitem{R90b}
Michel Raynal.
\newblock {\em Algorithmes du parallèlisme, le problème de l'exclusion
  mutuelle}.
\newblock Dunod informatique, 1990.

\bibitem{V00j}
George Varghese.
\newblock Self-stabilization by counter flushing.
\newblock {\em SIAM J. Comput.}, 30(2):486--510, 2000.

\end{thebibliography}
